\documentclass[a4paper,12pt]{article}

\usepackage{amsmath,amssymb}
\usepackage{bbm}
\usepackage{graphicx}
\usepackage[bookmarks=true]{hyperref}

\ifx\hypersetup\sadfkjashdfkxja\else
\hypersetup{pdftitle={}}
\hypersetup{pdfauthor={}}
\hypersetup{pdfsubject={}}
\hypersetup{pdfkeywords={}}
\hypersetup{plainpages=false}
\hypersetup{bookmarksnumbered=true}
\hypersetup{pdfstartview=FitH}
\hypersetup{pdfpagemode=UseNone}
\hypersetup{colorlinks=false}
\hypersetup{citebordercolor={.5 1 .5}}
\hypersetup{urlbordercolor={.5 1 1}}
\hypersetup{linkbordercolor={1 .7 .7}}
\fi

\numberwithin{equation}{section}

\usepackage[a4paper,left=2.8cm,right=2.8cm,top=2.8cm,bottom=2.8cm]{geometry}

\allowdisplaybreaks[3]

\usepackage[font={small},labelfont={bf},width=0.85\textwidth]{caption}

\makeatletter
\let\old@startsection=\@startsection
\renewcommand{\@startsection}[6]{\old@startsection{#1}{#2}{#3}{#4}{#5}{#6\mathversion{bold}}}
\makeatother

\def\[{\begin{equation}}
\def\]{\end{equation}}
\def\<{\begin{eqnarray}}
\def\>{\end{eqnarray}}

\ifx\genfrac\sdflkaj

\else

\fi

\newcommand{\Tr}{\mathop{\mathrm{Tr}}}

\let\oldPhi=\Phi
\let\oldPsi=\Psi
\let\oldGamma=\Gamma
\let\oldDelta=\Delta
\let\oldSigma=\Sigma
\let\oldLambda=\Lambda
\let\oldTheta=\Theta
\let\oldPi=\Pi
\renewcommand{\Phi}{\mathnormal{\oldPhi}}
\renewcommand{\Psi}{\mathnormal{\oldPsi}}
\renewcommand{\Gamma}{\mathnormal{\oldGamma}}
\renewcommand{\Sigma}{\mathnormal{\oldSigma}}
\renewcommand{\Delta}{\mathnormal{\oldDelta}}
\renewcommand{\Theta}{\mathnormal{\oldTheta}}
\renewcommand{\Lambda}{\mathnormal{\oldLambda}}
\renewcommand{\Pi}{\mathnormal{\oldPi}}

\newcommand{\matr}[2]{\left(\begin{array}{#1}#2\end{array}\right)}

\newcommand{\tprods}[2]{\langle#1#2\rangle}

\def\ni{\noindent}
\def\be{\begin{equation}}
\def\ee{\end{equation}}
\def\bsp{\be\begin{split}}
\def\la{\langle}
\def\ra{\rangle}
\def\dag{\dagger}
\def\wt{\widetilde}

\def\lr{\leftrightarrow}

\def\G{\Gamma}

\def\a{\alpha}
\def\b{\beta}
\def\g{\gamma}

\def\d{\delta}
\def\e{\epsilon}
\def\m{\mu}

\def\n{\nu}
\def\s{\sigma}
\def\r{\rho}
\def\l{\lambda}

\def\p{\partial}

\def\lr{\leftrightarrow}

\ifx\href\asklfhas\newcommand{\href}[2]{#2}\fi

\makeatletter
\def\mr@ignsp#1 {\ifx\:#1\@empty\else #1\expandafter\mr@ignsp\fi}%
\newcommand{\multiref}[1]{\begingroup
\xdef\mr@no@sparg{\expandafter\mr@ignsp#1 \: }%
\def\mr@comma{}%
\@for\mr@refs:=\mr@no@sparg\do{\mr@comma\def\mr@comma{,}\ref{\mr@refs}}%
\endgroup}
\makeatother

\newcommand{\hypref}[2]{\ifx\href\asklfhas #2\else\href{#1}{#2}\fi}

\renewcommand{\eqref}[1]{(\multiref{#1})}

\makeatletter
\newlength{\apb@width}
\newcommand{\autoparbox}[2][c]{\settowidth{\apb@width}{#2}\parbox[#1]{\apb@width}{#2}}

\makeatother

\renewcommand{\title}[1]{\vbox{\center\LARGE{#1}}\vspace{5mm}}
\renewcommand{\author}[1]{\vbox{\center\large{#1}}\vspace{5mm}}
\newcommand{\address}[1]{\vbox{\center\em#1}}
\newcommand{\email}[1]{\vbox{\center\tt#1}\vspace{5mm}}

\begin{document}
\bibliographystyle{utphys}

\newpage
\setcounter{page}{1}
\pagenumbering{arabic}
\renewcommand{\thefootnote}{\arabic{footnote}}
\setcounter{footnote}{0}

\begin{titlepage}
\title{\vspace{1.0in} {\bf Manifest $SO({\cal N})$ invariance and
    $S$-matrices of three-dimensional ${\cal N}=2,4,8$ SYM}}

\author{Abhishek Agarwal$^1$ and Donovan Young$^2$}

\address{$^1$American Physical Society\\
  1 Research Road\\
  Ridge, NY 11961, USA\\
  \vspace{.5cm}
  $^2$Niels Bohr Institute, Blegdamsvej 17, DK-2100 Copenhagen, Denmark}

\email{$^1$abhishek@ridge.aps.org, $^2$dyoung@nbi.dk}

\abstract{An on-shell formalism for the computation of $S$-matrices of
  SYM theories in three spacetime dimensions is presented. The
  framework is a generalization of the spinor-helicity formalism in
  four dimensions. The formalism is applied to establish the manifest
  $SO(\mathcal{N})$ covariance of the on-shell superalgebra relevant
  to $\mathcal{N} =2,4$ and $8$ SYM theories in $d=3$. The results are
  then used to argue for the $SO(\mathcal{N})$ invariance of the
  $S$-matrices of these theories: a claim which is proved explicitly
  for the four-particle scattering amplitudes. Recursion relations relating tree
  amplitudes of three-dimensional SYM theories are shown to follow
  from their four-dimensional counterparts. The results for the
  four-particle amplitudes are verified by tree-level perturbative
  computations and a unitarity based construction of the integrand
  corresponding to the leading perturbative correction is also
  presented for the ${\cal N}=8$ theory. For $\mathcal{N}=8$ SYM, the
  manifest $SO(8)$ symmetry is used to develop a map between the
  color-ordered amplitudes of the SYM and superconformal Chern-Simons
  theories, providing a direct connection between on-shell observables
  of D2 and M2-brane theories.  }

\end{titlepage}

\tableofcontents

\section{Introduction and summary}

Recent developments have uncovered a wealth of algebraic structures
and dualities pertaining to the scattering matrices of supersymmetric
Yang-Mills (SYM) theories in $3+1$ dimensions. In the case of
$\mathcal{N}=4$ SYM in $d=4$, the possibility that the particularly
elegant Parke-Taylor form of the maximal helicity violating (MHV)
amplitudes \cite{Parke:1986gb} is suggestive of deeper symmetries in
the theory had been appreciated for a long time \cite{Nair:1988bq,
Kosower:1987ic}. It is only in recent times that it has become clear
that the holomorphic form of the MHV amplitudes is a particular
manifestation of Yangian, dual-conformal and dual-superconformal
symmetries of the $S$-matrix of the gauge theory \cite{Witten:2003nn,
Drummond:2008cr, Drummond:2009fd, Drummond:2008vq, ArkaniHamed:2010kv}
- symmetries which are not manifest at the level of the Lagrangian of
the theory (for recent reviews on the subject
see \cite{sreview1,sreview2,sreview3,areview1,areview2}).  These
symmetries also manifest themselves in the form of recursion relations
that relate amplitudes with a given number of external legs to
products of amplitudes with lower number of external legs at a fixed
order in perturbation theory \cite{bcfw}.  Given that the usual Feynman
diagrammatic techniques involve thousands of Feynman diagrams even for
$\sim 10$ legs at tree-level, the algebraic relations are a great boon
even from a strictly pragmatic computational point of
view. Furthermore, as far as loop corrections are concerned, the
method of generalized unitarity allows one to compute higher order
corrections to a given amplitude using only the minimal on-shell
information, providing a much sought-after alternative to the
exponential computational complexity encountered by the standard
Feynman diagrammatic approach to loop corrections \cite{unitarity1,
unitarity2, unitarity3}. All these structures, insights and symmetries
have contributed to revealing the underlying analytic and algebraic
structures of the $S$-matrix of the four-dimensional superconformal
theory, along with compelling proposals for its form to all orders in
perturbation theory \cite{ArkaniHamed:2010kv, Mason-twistor}.

Given this enormous progress in our understanding of the $S$-matrix of
$\mathcal{N}=4$ SYM it is natural to ask if $S$-matrices of
non-conformal gauge theories continue to exhibit ``hidden'' structures
not captured by their Lagrangians as well. Of particular interest in
this regard are SYM theories in $d=2+1$. In three dimensions, theories
with $\mathcal{N}=2,4,$ and 8 supersymmetry inherit all the Poincar\'e
supersymmetries of the four-dimensional theories of which they are
dimensional reductions. However, the appearance of a dimensionful
coupling constant prevents them from being conformally invariant, even
classically. These theories thus provide a controlled departure from
the regime of classically conformally invariant four-dimensional
Yang-Mills theories whose scattering matrices have been explored in
the greatest detail following the developments alluded to
above\footnote{For recent progress in various aspects of highly
supersymmetric three-dimensional Yang-Mills theories
see \cite{aady1,aady2}}. Of special interest is the D2-brane
worldvolume theory described by the dimensional reduction of
$\mathcal{N}=4$ SYM to the three-dimensional sixteen-supercharge
non-conformal $\mathcal{N}=8$ SYM theory. The duality between D2 and
M2-brane theories imply that the $g^2_{YM} \rightarrow \infty$ limit
of this theory is described by a superconformal Chern-Simons
theory \cite{blg1,blg2,blg3,blg4,abjm} whose scattering matrix has been
shown to exhibit Yangian symmetries and other twistorial properties
that are very reminiscent of the $S$-matrix of $\mathcal{N}=4$
SYM \cite{scs1, scs2, scs3, scs4}. Thus, it might be expected that the
$S$-matrix of the three-dimensional gauge theory might contain special
structures that its Lagrangian obscures.

On the face of it, a direct connection between the $S$-matrices of the
two $\mathcal{N}=8$ three-dimensional gauge theories presents a
challenge. In the case of the gauge group being $SU(2)$,
the M2-brane theory has a manifest $SO(8)$ R-symmetry, which is
reflected in its $S$-matrix. More generally, it may be expected - and
it was indeed verified to all loop orders in the case of four-particle
amplitudes in \cite{scs1} - that the scattering matrices of
$\mathcal{N}\geq 4$ SCS theories reflect the global R-Symmetries of
their Lagrangians. SYM theories with $\mathcal{N}= 2,4$ or 8
supersymmetries, only posses $SO(\mathcal{N}-1)$ global $R$-symmetry
in their Lagrangians. Thus, for there to be any meaningful comparison
of the D2 and M2-brane scattering matrices it is imperative that we
understand how the extra $U(1)$ symmetry emerges in the Yang-Mills
theories. The situation clearly requires the development of on-shell
techniques for SYM theories that parallel the recent studies with M2
brane theories.

If one is to look at techniques used to study on-shell
properties of four-dimensional gauge theories as models for developing
$d=2+1$ on-shell methods, several ostensible arguments can be made to
suggest that four-dimensional SYM techniques do not readily adapt to a
three-dimensional context. The fundamental building block of $d=4$
on-shell techniques is the spinor-helicity basis for the vector and
spinor degrees of freedom, which is at the heart of many twistorial
aspects of the $S$-matrices of $d=4$ SYM theories. Thus, the absence
of a helicity degree of freedom in three dimensions appears to
present an immediate stumbling block. Furthermore, the absence of
conformal symmetries in the SYM theories of interest to this work
appears to rule out dual-conformal and dual-superconformal symmetries
-- symmetries that were a sufficient condition for the existence of 
infinite dimensional Yangian symmetries for the $S$-matrix of
$\mathcal{N}=4$ SYM in $d=4$ \cite{Drummond:2009fd} -- in d=3.

Given the motivations and technical caveats discussed above, we
develop a manifestly three-dimensional on-shell formalism for SYM
theories in this paper. Motivated by the four-dimensional
spinor-helicity framework, we use solutions of the massless Dirac
equation to construct gluon polarization vectors.  One of the
advantages of our formalism is that it allows us to explicitly track
the fate of the helicity degree of freedom when four-dimensional gauge
theories are reduced to $d=2+1$. We find that helicity is augmented to
a continuous $U(1)$ symmetry of the $S$-matrices of three-dimensional
SYM theories. Furthermore, this $U(1)$ degree of freedom couples to
the $SO(\mathcal{N}-1)$ R-symmetry of $\mathcal{N}=2,4$ and $8$
$d=2+1$ SYM theories to make the on-shell representation of the
supersymmetry algebra {\it manifestly} $SO(\mathcal{N})$ covariant. Put
differently, our formalism makes it transparent that the scalar
corresponding to the on-shell gluon in three dimensions couples to the
remaining $\mathcal{N}-1$ scalars of the gauge theories in a way that
makes the on-shell supersymmetry algebra the same as the (off-shell)
algebra for a free theory of $\mathcal{N}$ massless real scalars and
fermions. This symmetry enhancement has not been known to be manifest
in the Lagrangian except in the abelian limit where the gauge field
can be dualized into a scalar \cite{seiberg16}. What we exhibit in the paper
is the same phenomenon; but in the context of the $S$-matrices of the
corresponding non-abelian gauge theories.

Using our formalism we are able to show that the four-particle amplitudes
of $\mathcal{N} = 2,4,8$ $d=2+1$ SYM theories have a manifest
$SO(\mathcal{N})$ invariance to all orders in perturbation
theory. We also argue, though we do not provide a formal proof in this paper, that the  manifest
$SO(\mathcal{N})$ invariance should extend to the higher-particle amplitudes as well.
The four-particle amplitudes have the form\footnote{$s$ and $t$ are
  the standard Mandelstam variables, while $g_{YM}$ is the coupling
  constant of the Yang-Mills theory under consideration.}
\[
S_{YM}(t,s,g^2_{YM})S_{ijkl}(\{\mathcal{W}\}; t,s)\label{univ4}.
\]
where the ``universal'' term $S_{ijkl}(\{\mathcal{W}\}; t,s)$ only
depends on the species of particles being scattered $\{\mathcal{W}\}$
and contains all the $SO(\mathcal{N})$ dependence (in the indices
$i,j,k$ and $l$). We show that $S_{ijkl}(\{\mathcal{W}\}; t,s)$ is an
$SO(\mathcal{N})$ invariant tensor for all the theories under
consideration in this paper. Furthermore, for the special case of the
$\mathcal{N}=8$ theory with gauge group $SU(2)$, we show that
$S_{ijkl}(\{\mathcal{W}\}; t,s)$ is the same as the corresponding
quantity for the BLG theory computed in \cite{scs1}. This
analysis allows us to see a direct connection between an $SO(8)$
invariant physical observable in the D2 and M2-brane theories and
reduce the problem of establishing the duality (at least in the
context of four-particle amplitudes) to the asymptotic behavior of a
single function $S_{YM}(t,s,g^2_{YM})$ as $g^2_{YM} \rightarrow
\infty$.

With the general aim of studying $S$-matrix elements of various
three-dimensional theories in perturbation theory we also provide a
one-to-one map between $S$-matrices of three-dimensional gauge
theories and the corresponding matrices of the four-dimensional
theories that they are dimensional reductions of. In particular we
give a useful presentation of three and four-dimensional gamma
matrices and spinor-helicity bases in which all tree-level amplitudes
in three-dimensional SYM theories are obtained simply by setting the
fourth component of the momentum to zero in the known results for
four-dimensional amplitudes. We use the relations between three and
four-dimensional amplitudes to derive the analogs of the BCFW
relations for three-dimensional SYM theories. We also verify these
results by explicit perturbative calculations for the $d=2+1$ theories
of interest. In this context, we are also able to interpret the known
kinematical results in four dimensions such as the vanishing of all
helicity ``plus'' amplitudes as BPS conditions in three
dimensions. Finally, using the ${\cal N}=8$ theory as an illustrative
example, we are able to show that the integrands corresponding to loop
corrections for amplitudes in the three-dimensional theories (obtained
using generalized unitarity) are also obtained from known integrands
for the associated four-dimensional theories.

The paper is organized as follows. In section \ref{sec:onshell} we
discuss the on-shell SUSY algebra, concentrating on the ${\cal N}=2$
theory. We show that the algebra, and therefore the $S$-matrix,
manifests $SO(2)$ symmetry. The argument readily generalizes to
$SO(\mathcal{N})$ symmetry for theories with $\mathcal{N}>2$. We also
give an explicit map of the $d=4$ spinor-helicity formalism under
dimensional reduction to $d=3$ and obtain recursion relations for all
tree-level amplitudes. We derive specific relations between
four-particle scattering amplitudes following from the SUSY algebra. In
section \ref{sec:tree} we calculate all four-particle scattering
amplitudes in the ${\cal N}=2,4,8$ theories at tree-level directly
using Feynman diagrams. The manifest $SO({\cal N})$ forms of the
amplitudes are presented, which allows a verification of the relations
derived from the SUSY algebra in section \ref{sec:onshell}. In section
\ref{sec:oneloop} we discuss the four-particle amplitudes in the
${\cal N}=8$ theory at one-loop, recovering the scalar box integral
found in ${\cal N}=4$ SYM in $d=4$. In section \ref{sec:scs} we
discuss the relationship between the $S$-matrix of superconformal
Chern-Simons theories and the SYM theories considered in the paper. We
conclude the paper with a discussion in section \ref{sec:conc}.

\section{On-shell $\mathcal{N}=2$ algebra in three dimensions}
\label{sec:onshell}

In this section we obtain the $SO(\mathcal{N}) $ symmetric realization
of the on-shell supersymmetry algebra and introduce the
spinor-``helicity'' techniques tailored for the analyses of $S$-matrices
of $d=3$ gauge theories.  We then use our particular realization of
the algebra to constrain four-particle amplitudes to a single function of
the Mandelstam variables and $g^2_{YM}$, thereby fixing the ``matrix''
or R-symmetry structure of the four particle $S$-matrix. The
relationships so obtained between the various matrix elements are
later verified explicitly at weak coupling. We also present an
explicit map between the three-dimensional formalism developed in this
paper and the well known spinor-helicity framework in four
dimensions. As a byproduct of this map, we are able to relate $all$
tree-level three-dimensional amplitudes for $\mathcal{N}=2,4$ and $8$
theories to known results for $\mathcal{N}=1,2$ and $4$ theories in
one higher dimension very transparently. In the interests of brevity,
most of the discussion concerning our formalism will be limited to the
$d=3$, $\mathcal{N}=2$ case. The generalizations of the results
presented in this section to higher supersymmetry are straightforward
and we shall present the results relevant to higher extended
supersymmetry later in the paper.

The on-shell algebra acting on asymptotic scattering states is nothing
but the supersymmetry algebra of the free theory.  The free
$\mathcal{N}=2$ SYM theory can be expressed in a manifestly $SO(2)$
symmetric form by dualizing the gauge field to a scalar $\partial_\mu
\Phi_1 \sim \epsilon_{\mu \nu \rho} F^{\nu \rho}$. Note that since the
free theory is the same as the abelian limit, we can ignore color
indices for this discussion.  The dualized action 
\be\label{scalar}
S_d = \int d^3x\,\Bigl( -\frac{1}{2}\partial_\mu \Phi_I \partial^\mu \Phi_I +
\frac{i}{2}\bar{\chi }_I\gamma_\mu \partial ^\mu \chi_I\Bigr), \ee is
invariant under the $\mathcal{N} = 2$ SUSY transformations \bsp &
\delta \chi_I = - \frac{1}{2} \left(\partial_\mu \Phi_1 \gamma^\mu
\epsilon_I + \epsilon_{IJ} \partial_\mu \Phi_2 \gamma^\mu
\epsilon_J\right),\\ & \delta \Phi_1 = \frac{i}{2}\bar{\chi}_I
\epsilon_I,\\ & \delta \Phi_2 = \frac{i}{2}\bar{\chi_I}\epsilon_J
\epsilon_{IJ},
\end{split}
\ee
where $\epsilon_{12} = -\epsilon_{21} = +1$. Also, in the above
equations, $\delta \mathcal{W} = [\bar{\epsilon}_I Q_I,  \mathcal{W}]$
and we can read-off the algebra
\be [\bar{\beta} _M Q_M, \bar{\epsilon}_N Q_N] = \frac{1}{2}(
\bar{\epsilon}_L\gamma^\mu \beta_L )p_\mu, \ee
or equivalently
\be \{Q^\alpha_a, Q^\beta_b\} = \frac{1}{2}P^{\alpha
\beta}\delta_{ab}, \hspace{.2cm}\mbox{where} \hspace{.2cm} P^{\alpha
\beta} = - (p_\mu \gamma^\mu C^{-1})^{\beta \alpha}.\label{n2} 
\ee 
In
three dimensions, we can always pick a real Majorana representation
for the $\gamma $ matrices $\gamma ^\mu = (i\sigma ^2, \sigma ^1,
\sigma^3)$ with $C = \gamma ^0$, so that \be P^{\alpha \beta} =
P^{\beta \alpha } = \matr{cc}{-p_0 - p_1&p_2\\p_2&-p_0+p_1}.  \ee The
solution of the Dirac equation $\gamma ^\mu p_\mu u(p)=0$ is given by
\be u(p) = \frac{1}{\sqrt{p_0 - p_1}}\matr{c}{p_2 \\ p_1 - p_0},
\hspace{.2cm} u^\alpha (p) u^\beta(p) = -P^{\alpha \beta}.  \ee The
on-shell (momentum space) version of the $\mathcal{N}=2$ algebra can
be expressed as follows (we use $a $ and $\lambda$ to denote the
momentum-space creation operators for the $\Phi$ and $\chi$ fields respectively) \bsp &Q^\alpha _I |a_1\rangle =
\frac{1}{2} u^\alpha |\lambda_I\rangle,\\ &Q^\alpha _I |a_2\rangle =
\frac{1}{2}\epsilon_{IJ}u^\alpha |\lambda_J\rangle,\\ & Q^\alpha _J |
\lambda_I \rangle = -\frac{1}{2}u^\alpha \left(\delta_{JI}|a_1\rangle
+ \epsilon_{JI}|a_2\rangle\right).
\end{split}\label{n2so2}
\ee
This is the manifestly $SO(2)$ covariant form of the algebra which
should be realized on the asymptotic states of the gauge theory.

\subsection{Explicit realization}

To concretely establish that (\ref{n2so2}) is indeed the on-shell
representation of the SUSY algebra for $\mathcal{N}=2$ SYM in $d=2+1$,
in this section we will obtain the result via dimensional reduction,
starting with ${\cal N}=1$ SYM in $d=4$, whose action is given in (\ref{d4N4act}).
We start with a four-dimensional real representation of the $\Gamma$
matrices 
\bsp &\Gamma^M =
\Biggl\{\matr{cc}{i\sigma^2&0\\0&-i\sigma^2},
\matr{cc}{\sigma^1&0\\0&-\sigma^1},
\matr{cc}{\sigma^3&0\\0&-\sigma^3},
\matr{cc}{0&\mathbbm{1}\\\mathbbm{1}& 0}\Biggr\}, \\ &\Gamma^5 =
i\Gamma^0\cdots \Gamma^3.\label{4gamma}
\end{split}
\ee 
The four-dimensional Majorana fermion $\Psi = \begin{pmatrix}
\lambda_1\\ \lambda_2 \end{pmatrix}$. The three-dimensional $\gamma $
matrices are \be \gamma^\mu = \left(i\sigma^2, \sigma ^1,
\sigma^3\right).  \ee The dimensional reduction is carried out by
compactifying the ``3'' direction. It is readily seen that the fermion
kinetic energy term $\int \bar \Psi \Gamma^M\partial_M\Psi = \int
\bar{\lambda}_I\gamma^\mu\partial_\mu \lambda_I$ upon dimensional
reduction, leading us to identify $\lambda_I$ as the three-dimensional
fermions. It is understood that the charge conjugation matrix is
identified with $\Gamma^0$ and $\gamma^0$ in four and three dimensions
respectively and the Majorana conditions $\Psi^\dagger = \Psi^T$ and
$\lambda_I^\dagger = \lambda_I^T$ are imposed.  The four-dimensional
SUSY transformation law $\delta A_M = \frac{1}{2} \bar{\epsilon}
\Gamma_M \Psi$ (we will not display the color indices in this
subsection to avoid notational clutter) translates into the three-dimensional relation 
\bsp &\delta A_\mu = \frac{1}{2}
\bar{\eta}_I\gamma_\mu \lambda_I,\\ &\delta \Phi =
\frac{1}{2}\epsilon_{IJ}\bar{\eta}_I\lambda_J,
\end{split}\label{n2conventional}
\ee
where $A_3 = \Phi$.  We now want to translate these relations into
relations between momentum space physical degrees of freedom and
recover (\ref{n2so2}). For this purpose, it is very convenient to
introduce a polarization vector
\be \epsilon_\mu (p,k) =
\frac{\langle p|\gamma_\mu|k\rangle}{\tprods{k}{p}},\hspace{.3cm}
p_\mu \epsilon^\mu (p,k) = k_\mu \epsilon^\mu (p,k) = 0. 
\ee
It is
implied that
 \be |p\rangle = u(p), \hspace{.2cm} \langle p| =
\bar{u}(p), \hspace{.2cm} \tprods{k}{p} = \bar{u}(k)u(p) = -
\tprods{p}{k}, \ee
 where $u(p)$ is the  wavefunction defined before. The
polarization vectors satisfy
 \be\label{fi} \epsilon_\mu (p,k)
\epsilon^\mu (p,k') = +1, \hspace{.3cm} \epsilon_\mu (p,k)
(\gamma^\mu)_{\alpha \beta} = \frac{2 \bar{u}_\beta(p)u_\alpha(k) -
\d_{\a\b}\tprods{p}{k}}{\tprods{k}{p}} .
\ee
 To get the second equation above, we have used the Fierz identity:
$(\gamma_\mu)_{\alpha \beta}(\gamma^\mu)_{\gamma \delta} =
2\delta_{\alpha \delta}\delta_{\beta \gamma} - \delta_{\alpha
\beta}\delta_{\gamma\delta}$.

The dynamical fields are mode-expanded as follows,
\bsp \label{modeexp}
&\Phi =
\int \frac{d^2p}{(2\pi)^2}\frac{1}{\sqrt{2p^0}}\left(a_2^\dagger(p)e^{ip.x} +
a_2(p)e^{-ip.x}\right),\\ &A_\mu = \int
\frac{d^2p}{(2\pi)^2}\frac{1}{\sqrt{2p^0}}\epsilon_{\mu}(p,k)\left(a_1^\dagger(p)e^{ip.x}
+ a_1(p)e^{-ip.x} \right),\\ 
& \lambda_I= \int
\frac{d^2p}{(2\pi)^2}\frac{1}{\sqrt{2p^0}}\left(u(p) \lambda_I^\dagger(p)e^{ip.x} +
u(p)\lambda_I(p)e^{-ip.x} \right).
\end{split}
\ee
Application of (\ref{n2conventional}) to the oscillator expansion
given above implies that the on-shell states $a^\dagger_I|0\rangle =
|a_I\rangle$ transform as
\bsp\label{Q1}
&Q^\alpha _I |a_1\rangle = \frac{1}{2} u^\alpha |\lambda_I\rangle,\\
&Q^\alpha _I |a_2\rangle  = \frac{1}{2}\epsilon_{IJ}u^\alpha |\lambda_J\rangle,\\
\end{split}
\ee which are the first two equations of (\ref{n2so2}). The action for
the supercharge on the fermion field \be\label{Q2} Q^\alpha _J |
\lambda_I \rangle = -\frac{1}{2}u^\alpha \left(\delta_{IJ}|a_1\rangle
+ \epsilon_{JI}|a_2\rangle\right), \ee follows simply from the
condition of the closure of the algebra (\ref{n2}). We have thus
recovered the manifestly $SO(2)$ symmetric form of the on-shell
$\mathcal{N}=2$ algebra directly from the canonical quantization of
the gauge theory.  The $S$-matrix
\[
\langle \mathcal{W}_{I_1} \mathcal{W}_{I_2} \cdots \mathcal{W}_{I_n}\rangle = S_{I_1I_2\cdots I_n},
\]   
where $\mathcal{W}_J $ stand for any of the four bosonic or fermionic
fields must be $SO(2)$ invariant, since the component fields have a
manifest $SO(2)$ covariance and the $S$-matrix commutes with the
supercharges given above. We generalize these arguments in section
\ref{n4} to higher extended supersymmetry algebras.

\subsection{Recovering helicity}
\label{sec:rech}

Equivalently, one can recast the algebra in terms of a $U(1)$
symmetric form, which is very instructive for the purposes of drawing
parallels with known results for $\mathcal{N}=1$ SYM in $d=4$. Defining
the complex combinations $\mathcal{W}_\pm =
\frac{1}{\sqrt{2}}(\mathcal{W}_1 \pm i \mathcal{W}_2)$, we can express
the algebra on single particle states as
\bsp\label{Q3} 
&Q^\alpha _+ |a_+\rangle = \frac{1}{\sqrt{2}} u^\alpha
|\lambda_+\rangle, \hspace{.4cm}Q^\alpha _+ |\lambda_-\rangle = -
\frac{1}{\sqrt{2}} u^\alpha |a_-\rangle,\\ &Q^\alpha _- |a_-\rangle =
\frac{1}{\sqrt{2}} u^\alpha |\lambda_-\rangle, \hspace{.4cm}Q^\alpha
_- |\lambda_+\rangle = - \frac{1}{\sqrt{2}} u^\alpha |a_+\rangle,\\
&Q^\alpha _- |a_+\rangle = Q^\alpha _+ |a_-\rangle = Q^\alpha _+
|\lambda_+\rangle = Q^\alpha _- |\lambda_-\rangle = 0.
\end{split}
\ee
The $a_\pm$ ($\l_\pm$) states are direct analogues of the $d=4$ gluon
(gluino) helicities. In the next subsection we consider the fate of the
helicity degree of freedom under dimensional reduction from $d=4$ in detail.

\subsection{Dimensional reduction of the spinor-helicity basis from
  $d=4$ to $d=3$}
\label{dimredhelicity}

The on-shell supersymmetry transformations given above were derived in
a purely three-dimensional set up. We now study how the above
discussion is related to the spinor-helicity framework in four
dimensions. With the four-dimensional gamma matrices as in
(\ref{4gamma}), eigenstates of the helicity operator $\Gamma _\pm =
\frac{1}{2}(1 \pm \Gamma^5)$ are given by spinors of the form
\bsp &U_+ = \frac{1}{\sqrt{2}} \begin{pmatrix} V\\ -iV \end{pmatrix},
\hspace{.5cm}\mbox{where}\hspace{.5cm} p_\mu\gamma^\mu V = +ip_3V,\\
&U_- = \frac{1}{\sqrt{2}} \begin{pmatrix} U\\ iU \end{pmatrix},
\hspace{.5cm}\mbox{where}\hspace{.5cm} p_\mu\gamma^\mu U = -ip_3U.
\end{split}
\ee 
The four-dimensional massless Dirac equation, written on the right of
the above equations is readily interpreted as two copies of the massive
Dirac equation satisfied by $U$ and $V$, with masses $\sim \pm p_3$.
The solutions for $U$ and $V$ are
 \bsp V = \frac{1}{\sqrt{p_0 -
p_1}}\matr{c}{p_2 + ip_3\\ p_1 - p_0},\hspace{.5cm} U =
\frac{1}{\sqrt{p_0 - p_1}}\matr{c}{p_2 -ip_3\\ p_1 - p_0}.
\end{split}
\ee
We also have the closure\footnote{The solutions $U, V$ can also be
  regarded as wavefunctions for massive fermions in three dimensions,
  with $p_3\sim m$. Indeed, in \cite{scs1} these solutions were used
  extensively in the computations of four-particle amplitudes of massive
  SCS theories. } conditions $V^\alpha(p)U^\beta(p) = -P^{\alpha
  \beta} -ip_3\epsilon^{\alpha \beta}$. The two wavefunctions are
related by complex conjugation as $U^*(p) = -iV(-p)$. Most importantly
for our considerations $U(p)_{p_3=0} = V(p)_{p_3=0} = u(p)$, where the
momentum of the last wavefunction is three-dimensional. The
four-dimensional spinor products can be expressed as
\be \tprods{p}{q} = +\bar{U}(p)V(q), \hspace{.3cm} [pq] =
+\bar{V}(p)U(q), \hspace{.3cm} |\tprods{p}{q}|^2 = -2p\cdot q.  \ee
The four-dimensional polarization vectors are chosen in-line with
standard conventions
\[
\epsilon ^\pm _M (p,k) = \pm \frac{\tprods{p\pm|\Gamma _M}{|k\pm}}{\sqrt{2}\tprods{k\mp|}{p\pm}},
\]
where $k$ is a reference four-momentum, and decompose, upon
dimensional reduction to
\be \epsilon^{\pm \mu}(p,k) = +\frac{1}{\sqrt{2}}\epsilon^\mu (p,k),
\hspace{.3cm} \epsilon^{+3}(p,k) = -\frac{i}{\sqrt{2}}, \hspace{.3cm}
\epsilon^{-3}(p,k) = +\frac{i}{\sqrt{2}}.  \ee
The explicit split of the four-dimensional photon into a
three-dimensional photon and a scalar upon the compactfication of the
``3'' direction is as follows
\bsp A^M &= \int
\frac{d^2p}{(2\pi)^2}\frac{1}{\sqrt{2p^0}}\epsilon^{\pm M}(p,k)A^{\pm
\dagger}(p)e^{ip.x} + \text{h.c.}\\ & = \int
\frac{d^2p}{(2\pi)^2}\frac{1}{2\sqrt{p^0}}\left[\epsilon^{\pm
\mu}(p,k)\left(A^{+ \dagger}(p) + A^{- \dagger}(p) \right) -i
\left(A^{+ \dagger}(p) - A^{- \dagger}(p)\right)\right]e^{ip.x} +
\text{h.c.}.
\end{split}
\ee
We can thus readily identify
\bsp a_1^\dagger = \frac{1}{\sqrt{2}}(A^+ + A^-)^\dagger,
\hspace{.3cm} a_2^\dagger = -\frac{i}{\sqrt{2}}(A^+ - A^-)^\dagger,
\hspace{.3cm} a^{\pm \dagger} = (A^\pm)^\dagger.
\end{split}
\ee 
Consequently we can relate $all$ tree-level $n$-particle scalar/gluon
amplitudes as 
\be \langle++---+\cdots\rangle |_{d=3} =
\langle++---+\cdots\rangle|_{d=4, x_3=0}.  
\ee 
Similar formulae exist
for amplitudes involving fermions, namely 
\be |\lambda_+\rangle =
|\Psi_+\rangle_{x_3=0},\hspace{.3cm} |\lambda_-\rangle =
|\Psi_-\rangle_{x_3=0}.  
\ee 
Finally, if the field theory contains
scalar degrees of freedom $\Phi$, then one trivially has 
\be
|\Phi\rangle_{d=4, x_3 = 0}=|\Phi\rangle_{d=3}, 
\ee 
as the identifying relation between the three-dimensional on-shell
state and the dimensional reduction of its four-dimensional
counterpart.

These identifications allow us to read-off all the $\mathcal{N}=2,4$
and $8$ $d=3$ tree-level amplitudes from the known results for
$\mathcal{N}=1,2$ and $4$ SYM in $d=4$. The spinor bases in three and
four dimensions chosen here enable the map between the three and
four-dimensional amplitudes to be as simple as possible. Given an
amplitude in the four-dimensional theory, one simply sets the fourth
component of the momentum to zero to read-off the three-dimensional
results.\\

\ni{\bf Recursion relations:} The map between the three and
four-dimensional amplitudes also allows us to readily derive
tree-level recursion relations for three-dimensional amplitudes. Using
the notation of \cite{bcfw} and given a four-dimensional amplitude
$\mathcal{A}$, we define the amplitude with the momenta shifted by a
complex parameter $z$, $\mathcal{A}(z)$.  Taking the limit where the
fourth components of all the momenta are set to zero, after subjecting
the shifted amplitude to the BCFW relations yields recursion relations
for the three-dimensional amplitudes. In other words,
\be
\lim_{k^3_i\rightarrow 0}\left(\frac{1}{2\pi i}\oint dz
\frac{\mathcal{A}(z)}{z} = \mathcal{A}(0) -
\sum_{ij}\sum_h\frac{\mathcal{A}^h_L(z_{ij})\mathcal{A}^{-h}_R(z_{ij})}{P^2_{ij}}\right),
\ee 
is the relevant recursion relation for SYM amplitudes in $d=3$. Thus
to find the recursion relations for a given three-dimensional
amplitude, we can ``oxidize'' it to its four-dimensional counterpart
using the dictionary given above, apply the BCFW relations and then
dimensionally reduce back to three dimensions to obtain the desired
recursion relations.

\subsection{Relations between amplitudes}

We can use the complex form of the algebra to relate various
amplitudes. We start out by noticing that all amplitudes of the form
$\langle a_+\cdots a_+\rangle$, $\langle \lambda_+\cdots
\lambda_+\rangle$ and their complex conjugates are annihilated by half
of the supercharges, making them $\frac{1}{2}$ BPS
states. Furthermore, they are identically zero to all orders in
perturbation theory. For example
\be
\langle [Q_-, \lambda_+a_+\cdots a_+]\rangle = 0 \Rightarrow \langle a_+\cdots a_+\rangle = 0.
\ee
These are then the analogues of all helicity ``plus'' amplitudes in the
four-dimensional case.  Similarly, acting with $Q_-$ 
allows us to see that all boson amplitudes with only one $a_-$ -- the
analogues of the minimally helicity violating amplitudes in four
dimensions -- are also zero. Turning our attention to four-particle
amplitudes, it is easy to see that the only non-vanishing amplitudes
are those that involve two plus and two minus fields. Furthermore,
the only non-vanishing mixed amplitudes are those that involve
bosons and fermions of both signs, i.e. of the type $\langle
a_+a_-\lambda_+\lambda_-\rangle$.  These amplitudes are related to the
four boson amplitudes as follows
\bsp\label{bbff} &\langle \lambda_+ \lambda_- a_+a_-\rangle =
+\frac{\tprods{3}{2}}{\tprods{3}{1}}\langle a_+a_-a_+a_-\rangle,\quad
\langle \lambda_+ \lambda_- a_-a_+\rangle =
+\frac{\tprods{4}{2}}{\tprods{4}{1}}\langle a_+a_-a_-a_+\rangle,\\
&\langle a_+\lambda_- \lambda_+ a_-\rangle =
-\frac{\tprods{1}{2}}{\tprods{1}{3}}\langle a_+a_-a_+a_-\rangle,\quad
\langle a_+\lambda_- a_-\lambda_+ \rangle =
-\frac{\tprods{1}{2}}{\tprods{1}{4}}\langle a_+a_-a_-a_+\rangle,\\
&\langle \lambda_+ a_+\lambda_-a_- \rangle =
+\frac{\tprods{2}{3}}{\tprods{2}{1}}\langle a_+a_+a_-a_-\rangle,\quad
\langle a_+ \lambda_+\lambda_-a_- \rangle =
+\frac{\tprods{1}{3}}{\tprods{1}{2}}\langle a_+a_+a_-a_-\rangle,\\
&\langle \lambda_+ a_-\lambda_- a_+\rangle =
+\frac{\tprods{4}{3}}{\tprods{4}{1}}\langle a_+a_-a_-a_+\rangle,\quad
\langle a_+a_-\lambda_- \lambda_+ \rangle =
-\frac{\tprods{1}{3}}{\tprods{1}{4}}\langle a_+a_-a_-a_+\rangle,\\
&\langle a_+a_- \lambda_+ \lambda_-\rangle =
+\frac{\tprods{1}{4}}{\tprods{1}{3}}\langle a_+a_-a_+a_-\rangle,\quad
\langle \lambda_+a_-a_+ \lambda_- \rangle =
+\frac{\tprods{3}{4}}{\tprods{3}{1}}\langle a_+a_-a_+a_-\rangle,\\
&\langle a_+\lambda_+a_- \lambda_- \rangle =
+\frac{\tprods{1}{4}}{\tprods{1}{2}}\langle a_+a_+a_-a_-\rangle,\quad
\langle \lambda_+a_+a_- \lambda_- \rangle =
-\frac{\tprods{3}{1}}{\tprods{3}{4}}\langle a_+a_+a_-a_-\rangle.\\
\end{split}
\ee
The rest of the two-boson two-fermion amplitudes are related to the
ones given above by complex conjugation. Proceeding to the
four-fermion amplitudes,  we find the following relations
\bsp\label{ffff} &\langle \lambda_+\lambda_+\lambda_-\lambda_-\rangle
= +\frac{\tprods{1}{2}}{\tprods{2}{4}}\langle
a_+\lambda_+\lambda_-a_-\rangle =
+\frac{\tprods{1}{3}}{\tprods{2}{4}}\langle a_+a_+a_-a_-\rangle ,\\
&\langle \lambda_+\lambda_-\lambda_+\lambda_-\rangle =
+\frac{\tprods{4}{2}}{\tprods{4}{1}}\langle
a_+a_-\lambda_+\lambda_-\rangle =
+\frac{\tprods{2}{4}}{\tprods{1}{3}}\langle a_+a_-a_+a_-\rangle,\\
&\langle \lambda_+\lambda_-\lambda_-\lambda_+\rangle =
+\frac{\tprods{1}{4}}{\tprods{4}{2}}\langle
a_+a_-\lambda_-\lambda_+\rangle =
+\frac{\tprods{1}{3}}{\tprods{2}{4}}\langle a_+a_-a_-a_+\rangle.
\end{split}
\ee
These relationships imply that there is only one independent
four-particle amplitude for the $\mathcal{N}=2$ theory in three
dimensions. The number of independent amplitudes stays the same for
theories with higher extended supersymmetries as well. All the above
mentioned relationships have been verified explicitly at tree-level,
see section \ref{sec:tree}.

\subsection{Higher extended on-shell supersymmetry algebras}\label{n4}

Here we outline how the methods presented above can be used to derive
the on-shell supersymmetry algebra for theories with higher extended
supersymmetry, using the $\mathcal{N}=4$ supersymmetric case as an
illustrative example. The ${\cal N}=8$ case follows similarly; we
have relegated the analogous details to the appendices. We start with
$\mathcal{N}=1$ SYM in $d=6$ with the action
\be\label{N1D6} S = \int d^6x\,\Bigl( -\frac{1}{4}F^a_{MN}F^{aMN} +
\frac{i}{2} \bar{\Psi}^a \Gamma_M'D^M\Psi^a\Bigr) ,
\ee 
invariant under
\bsp\label{D6SUSY} &\delta A^a_M = \frac{i}{2}(\bar{\Psi}^a\Gamma_M'
\e - \bar{\e}\Gamma_M'\Psi^a),\\ &\delta \Psi^a =
-\frac{1}{4}[\Gamma_M',\Gamma_N']F^{aMN} \e.
\end{split}
\ee
The six-dimensional gamma matrices are related to the four-dimensional
ones (\ref{4gamma}) as
\be \Gamma' = \sigma^1\otimes \Gamma, \hspace{.2cm}
\sigma^1\otimes\Gamma^5, \hspace{.2cm}
\sigma^2\otimes\mathbbm{1}.\label{6gamma} \ee
In six dimensions, one can only have a Weyl condition, $(1
+\Gamma '^0 \cdots \Gamma '^5)\Psi = 0$, which is satisfied by
\be
\Psi =
\begin{pmatrix} 0\\ \Lambda \end{pmatrix},
\ee
where $\Lambda $ is a
four-dimensional Dirac fermion, which can be decomposed into two real
Majorana fermions as $\Lambda = \Lambda^1 + i\Lambda^2$, where
${\Lambda ^{i}}^T = {\Lambda^i}^\dagger$. The Majorana fermions can be
further decomposed into four, three-dimensional Majorana fermions
$\lambda_A$ with $A=1,\ldots,4$, as 
\be
\Lambda^1 = \begin{pmatrix}
  \lambda_1\\ \lambda _3 \end{pmatrix},\qquad \Lambda^2
= \begin{pmatrix} \lambda_2\\ \lambda _4 \end{pmatrix}.
\ee
After dimensionally reducing the theory to
three dimensions and using the spinor formalism described above, we
can read-off the action of the supersymmetry generators on the
on-shell degrees of freedom from (\ref{D6SUSY}). The result is 
\bsp
Q_A^\a |a_B\ra = \frac{1}{2}u^\a\,\r^B_{AC}|\l_C\ra,\qquad
Q_A^\a |\l_B\ra = -\frac{1}{2} u^\a\,\r^C_{AB} |a_C\ra,
\end{split}
\ee
where $a_{A=1}$ represents the $d=3$ gluon and $a_{A\neq 1}$ the three
scalars of the theory, and where\footnote{The upper index on
  $\r^A_{BC}$ labels the elements in the list, so that the lower
  indices are the indices of the $4\times4$ matrices.} 
\be\label{n4rhos}
\r^{A}_{BC}=\Bigl\{ \mathbbm{1}\otimes\mathbbm{1},~ 
i\s^2\otimes\mathbbm{1},-\s^1\otimes i\s^2,~\s^3\otimes i \s^2 \Bigr\}.
\ee
On any of the bosonic or fermionic states $|\mathcal{W}\rangle$ the
algebra closes in an $SO(4)$ symmetric form as
\be \{(Q_A)^\alpha, (Q_B)^\beta\}|\mathcal{W}\rangle =
+\frac{1}{2} P^{\alpha \beta}\delta
_{AB}|\mathcal{W}\rangle.
\ee
In this form the $SO(4)$ covariance of the $\mathcal{N}=4$
supersymmetry algebra is manifest. As a matter of fact, it is readily
seen that this algebra is a symmetry of the $SO(4)$ invariant free
$\mathcal{N}=4$ action $S = \int -\frac{1}{2}\partial_\mu \Phi_A
\partial^\mu \Phi_A + \frac{i}{2}\bar{\lambda }_A\gamma_\mu
\partial ^\mu \lambda_A$.  Thus the four-particle scattering matrix
$\langle \mathcal{W}_A\mathcal{W}_B
\mathcal{W}_C\mathcal{W}_D\rangle $ must have the form
\be
 \langle \mathcal{W}_A\mathcal{W}_B
 \mathcal{W}_C\mathcal{W}_D\rangle = {\cal A}\,\delta_{AB}\delta
 _{CD} + {\cal B}\,\delta_{AC}\delta_{BD} + {\cal C}\,\delta
_{AD}\delta_{BC} + {\cal D}\,\epsilon_{ABCD}, 
\ee
 for it to commute with the supercharges given above. More
 generally, $n$-particle amplitudes of this gauge theory must only involve
 $SO(\mathcal{N})$ invariants for the $S$-matrix to commute with the
 on-shell supersymmetry generators.  The relations between the
 undetermined coefficients as well as the extension of the formalism
 to the case of $\mathcal{N}=8$ SUSY is discussed in the next chapter,
 where the perturbative results for $\mathcal{N}=2,4$, and $8$ theories
 are presented in a unified manner.

\section{Tree-level four particle amplitudes}
\label{sec:tree}

In order to have an explicit check of the relations put forth in the
preceding sections, in this section we will give results for
four-particle scattering in ${\cal N} = 2,4$ and ${\cal N}=8$ SYM at
tree-level, from a direct Feynman diagram calculation. Our conventions
are collected in appendix \ref{app:dimred}.

The SYM action may be expressed\footnote{The action for the ${\cal
    N}=2$ (${\cal N}=8$) theory is derived by dimensional reduction
    from the ${\cal N}=1$ theory in $d=4$ ($d=10$) in appendix
    \ref{app:dimred}. The ${\cal N}=4$ theory is discussed in section \ref{n4}.} (in mostly positive signature) as
\be
S = \frac{1}{g^2} \Tr \int d^3 x 
\left(-\frac{1}{2} F_{\m\n} F^{\m\n} -D_\m\Phi_i D^\m\Phi_i+ i \bar \l_A \g^\m D_\m \l_A 
+\r^i_{AB}\bar \l_A [\Phi_i,\l_B] 
\right),
\ee
where $\l_A$ are Majorana 2-spinors, with $A=1,\ldots,{\cal N}$, while
the $\Phi_i$ are real scalars with $i=2,\ldots,{\cal N}$. The Yukawa
couplings are given by $\r_{AB}^i = \e_{AB}$ for ${\cal N}=2$, by
(\ref{n4rhos}) for ${\cal N}=4$, and
for ${\cal N}=8$ they are given by the matrices relating the ${\bf
8}_v$, ${\bf 8}_c$, and ${\bf 8}_s$ representations of $SO(8)$, see
(\ref{rhos}). The set of such matrices is completed with the unit
matrix, so that $\r^C_{AB}=\{\r^1_{AB}=\d_{AB},\r^i_{AB}\}$.  All
fields transform in the adjoint representation of $SU(N)$.  The mode
expansions are a slight generalization of (\ref{modeexp})
\bsp \label{modeexp1}
&\Phi_i =
\int \frac{d^2p}{(2\pi)^2}\frac{1}{\sqrt{2p^0}}\left(a_i^\dagger(p)e^{ip.x} +
a_i(p)e^{-ip.x}\right),\\ &A_\mu = \int
\frac{d^2p}{(2\pi)^2}\frac{1}{\sqrt{2p^0}}\epsilon_{\mu}(p,k)\left(a_1^\dagger(p)e^{ip.x}
+ a_1(p)e^{-ip.x} \right),\\ 
& \lambda_A= \int
\frac{d^2p}{(2\pi)^2}\frac{1}{\sqrt{2p^0}}\left(u(p) \lambda_A^\dagger(p)e^{ip.x} +
u(p)\lambda_A(p)e^{-ip.x} \right).
\end{split}
\ee

We will be interested in calculating colour-ordered
amplitudes. Labelling the four particles' gauge group indices as $a_1$,
$a_2$, $a_3$, and $a_4$, one generically finds expressions depending upon the
following contractions of the gauge group structure constants
\bsp
&f^{a1 t a2} f^{a3 t a4} = -2 \Tr[T^{a_1}T^{a_2}T^{a_3}T^{a_4}] 
+2 \Tr[T^{a_2}T^{a_1}T^{a_3}T^{a_4}]\\&\qquad\qquad\qquad +2
\Tr[T^{a_3}T^{a_1}T^{a_2}T^{a_4}] 
-2 \Tr[T^{a_3}T^{a_2}T^{a_1}T^{a_4}],\\
&f^{a1 t a4} f^{a3 t a2} = -2 \Tr[T^{a_1}T^{a_2}T^{a_3}T^{a_4}] 
+2 \Tr[T^{a_4}T^{a_1}T^{a_3}T^{a_2}]\\&\qquad\qquad\qquad +2 \Tr[T^{a_3}T^{a_1}T^{a_4}T^{a_2}] 
-2 \Tr[T^{a_3}T^{a_2}T^{a_1}T^{a_4}],\\
&f^{a1 t a3} f^{a2 t a4} = -2 \Tr[T^{a_1}T^{a_3}T^{a_2}T^{a_4}] 
+2 \Tr[T^{a_3}T^{a_1}T^{a_2}T^{a_4}]\\&\qquad\qquad\qquad +2 \Tr[T^{a_2}T^{a_1}T^{a_3}T^{a_4}] 
-2 \Tr[T^{a_2}T^{a_3}T^{a_1}T^{a_4}].
\end{split}
\ee
The colour-ordered contributions are those proportional to
$\Tr[T^{a_1}T^{a_2}T^{a_3}T^{a_4}]$, and so come from $f^{a1 t a2}
f^{a3 t a4}$ and $f^{a1 t a4} f^{a3 t a2}$. In what follows
we have restored the more conventional counting of the coupling
constant by rescaling all fields by $g$. We find it most convenient to
present the colour-ordered amplitude at tree-level in the
following way\footnote{Note that the coupling $g^2$ should be
  understood to be made dimensionless via the introduction of a
  renormalization scale $\m$, so that $g^2 \sim g^2 /\m$.}
\be
\Bigl\la {\phi^{a_1}_{{\cal A}_1}}^\dag(p_1)\, {\phi^{a_2}_{{\cal A}_2}}^\dag(p_2)\, 
{\phi^{a_3}_{{\cal A}_3}}^\dag(p_3)\, {\phi^{a_4}_{{\cal A}_4}}^\dag(p_4)\, \Bigr\ra = 2ig^2\, {\cal
  C}\left(\phi_{{\cal A}_1}\phi_{{\cal A}_2} \phi_{{\cal A}_3}\phi_{{\cal
    A}_4} \right)\Tr[T^{a_1}T^{a_2}T^{a_3}T^{a_4}] + \ldots
\ee
where $\phi_{\cal A}^\dag(p) ={\phi^{a}_{\cal A}}^\dag(p)T^a $ is the creation
operator for the given field $\phi_{\cal A}$ as per the mode expansions given
in (\ref{modeexp1}), ${\cal A}$ denotes a general flavour index, and the
``$\ldots$'' refers to non-colour-ordered contributions. All momenta
are taken to be in-going, so that $\sum_{i=1}^4p_i^\m = 0$.

\subsection{Four fermion scattering}

There are two Feynman diagrams contributing to the scattering of four
$\l^\dag_A(p)$ external states; the gluon exchange and the scalar
exchange. We may express these two contributions in terms of the
following two expressions respectively
\bsp
&{\cal A}(1,2,3,4) \equiv \frac{\Bigl(\bar u(p_1) \g^\m u(p_2)\Bigr)
\Bigl(\bar u(p_3) \g_\m u(p_4)\Bigr)}{(p_1+p_2)^2},\\
&{\cal B}(1,2,3,4) \equiv \frac{\Bigl(\bar u(p_1) u(p_2)\Bigr)
\Bigl(\bar u(p_3) u(p_4)\Bigr)}{(p_1+p_2)^2}.
\end{split}
\ee 
Note that ${\cal A}(1,2,3,4)={\cal A}(2,1,3,4) ={\cal
  A}(1,2,4,3)={\cal A}(2,1,4,3)$, while ${\cal B}(1,2,3,4)=-{\cal B}(2,1,3,4) =-{\cal
  B}(1,2,4,3)={\cal B}(2,1,4,3)$. We find
\bsp
{\cal C}\left(\l_{A_1}\l_{A_2}\l_{A_3}\l_{A_4}\right)
=
& \d_{A_1A_2} \d_{A_3A_4} \Bigl( {\cal B}(4,1,2,3)+
 {\cal A}(1,2,3,4)
 \Bigr) \\
-& \d_{A_1A_3} \d_{A_2A_4} \Bigl(
{\cal B}(4,1,2,3) -  {\cal B}(1,2,3,4)\Bigr)\\
-& \d_{A_1A_4} \d_{A_2A_3}\Bigl( 
{\cal A}(4,1,2,3)
+  {\cal B}(1,2,3,4)\Bigr).
\end{split}
\ee
We note that the amplitude is manifestly $SO({\cal N})$ invariant.

\subsection{Two boson - two fermion tree-level amplitudes}

We begin by calculating the amplitudes
\bsp
&\Bigl\la {a^{a_1}_{i_1}}^\dag(p_1)\, {a^{a_2}_{i_2}}^\dag(p_2)\, 
{\l^{a_3}_{A_3}}^\dag(p_3)\, {\l^{a_4}_{A_4}}^\dag(p_4)\, \Bigr\ra,\\
&\Bigl\la {a_1^{a_1}}^\dag(p_1)\, {a^{a_2}_{i_2}}^\dag(p_2)\, 
{\l^{a_3}_{A_3}}^\dag(p_3)\, {\l^{a_4}_{A_4}}^\dag(p_4)\, \Bigr\ra,\\
&\Bigl\la {a^{a_1}_{i_1}}^\dag(p_1)\, {a_1^{a_2}}^\dag(p_2)\, 
{\l^{a_3}_{A_3}}^\dag(p_3)\, {\l^{a_4}_{A_4}}^\dag(p_4)\, \Bigr\ra,\\
&\Bigl\la {a_1^{a_1}}^\dag(p_1)\, {a_1^{a_2}}^\dag(p_2)\, 
{\l^{a_3}_{A_3}}^\dag(p_3)\, {\l^{a_4}_{A_4}}^\dag(p_4)\, \Bigr\ra.
\end{split}
\ee
There are contributions from a fermion exchange
\bsp
&{\cal C}_F\left(a_1\,a_1\,\l_{A_3}\l_{A_4}\right)
 = -\delta_{A_3A_4}\,
\bar u(p_4) {\not \e}(p_1) \frac{({\not p_2}+
    {\not p_3})}{(p_2+p_3)^2} {\not \e}(p_2)  u(p_3),\\
&{\cal C}_F\left(a_{i_1} a_{i_2} \l_{A_3}\l_{A_4}\right)
=\Bigl(\left(\r^{i_1}\right)^T \r^{i_2}\Bigr)_{A_4A_3} \,
\bar u(p_4) \frac{({\not p_2}+
    {\not p_3})}{(p_2+p_3)^2}  u(p_3),\\
&{\cal C}_F\left(a_{i_1} \,a_1\, \l_{A_3}\l_{A_4}\right)
 = \r^{i_1}_{A_3A_4} \,
\bar u(p_4) \frac{({\not p_2}+
    {\not p_3})}{(p_2+p_3)^2} {\not \e}(p_2)  u(p_3),\\
&{\cal C}_F\left(a_1\,a_{i_2} \l_{A_3}\l_{A_4}\right)
 = \r^{i_2}_{A_3A_4} \, 
\bar u(p_4) {\not \e}(p_1) \frac{({\not p_2}+
    {\not p_3})}{(p_2+p_3)^2}  u(p_3),
\end{split}
\ee
and a boson exchange,
\bsp
{\cal C}_B&\left(a_1\,a_1\,\l_{A_3}\l_{A_4}\right)
=-\delta_{A_3A_4}\, 
 \Biggl[ -2 p_1 \cdot \e(p_2) \frac{\bar u(p_4) {\not \e}(p_1)u(p_3)}{(p_1+p_2)^2} +
2 p_2 \cdot \e(p_1) \frac{\bar u(p_4) {\not \e}(p_2)u(p_3)}
{(p_1+p_2)^2}\\
&\qquad\qquad +\e(p_1)\cdot \e(p_2) \frac{\bar u(p_4) \left({\not p_1}-{\not
    p_2}\right)u(p_3)}
{(p_1+p_2)^2}  \Biggr],\\
{\cal C}_B&\left(a_{i_1} a_{i_2} \l_{A_3}\l_{A_4}\right)
=-\d_{i_1i_2}\delta_{A_3A_4} \, \frac{\bar u(p_4) \left({\not p_1}-{\not
    p_2}\right)u(p_3)}
{(p_1+p_2)^2},\\
{\cal C}_B&\left(a_{i_1} \,a_1\, \l_{A_3}\l_{A_4}\right)
 = -2\r^{i_1}_{AB} \, \left( p_1 \cdot \e(p_2) \right)\,
 \frac{\bar u(p_4)u(p_3)}{(p_1+p_2)^2} ,\\
{\cal C}_B&\left(a_1\,a_{i_2} \l_{A_3}\l_{A_4}\right)
 = 2\r^{i_2}_{AB} \, \left(p_2 \cdot \e(p_1)\right)\,
 \frac{\bar u(p_4)u(p_3)}{(p_1+p_2)^2}.
\end{split}
\ee
The complete tree-level amplitudes are obtained by taking the sum
of the boson and fermion exchanges. 

One may determine the remaining amplitudes as follows 
\bsp\label{rear}
&{\cal C}\left(a_D \l_A \l_B a_C \right) 
= {\cal C}\left(a_C a_D \l_A \l_B \right) ~\text{with}~ 
p_1 \to p_4,~p_2\to p_1,~ p_3\to p_2,~p_4 \to p_3,\\
&{\cal C}\left(a_C \l_A a_D \l_B \right) =
  -{\cal C}\left(a_C a_D \l_A \l_B \right) ~\text{with}~ 
p_2 \lr p_3 - {\cal C}\left(a_C \l_A \l_B a_D \right)  ~\text{with}~ 
p_3 \lr p_4,
\end{split}
\ee
where $a_C$ indicates either a scalar or a gauge field, see
(\ref{together}). The amplitudes with a fermion in the first position
may be determined through applications of similar rules starting with
\be\label{fermfirst}
{\cal C}(\l_D \l_C a_B a_A) = {\cal C}(a_A a_B \l_C \l_D )
~\text{with}~ p_1 \lr p_4,~p_2 \lr p_3.
\ee 

\subsection{Four boson tree-level amplitudes}

The four boson amplitude stems from a boson exchange
diagram and a contact diagram stemming from the 4-boson
vertices. We find that the results may be compactly expressed by enlarging the
index $i$ on the scalar field to include a first component
which is identified with the gauge field degree of freedom,
i.e.
\be\label{together}
a_A^\dag = \left(a_1^\dag,a_i^\dag\right).
\ee
The color-ordered amplitude is then read-off from the following
compact expression
\bsp
{\cal C}\left(a_{A_1}a_{A_2}a_{A_3}a_{A_4}\right)\to
&\frac{1}{(p_1+p_2)^2}
\Biggl[ \Theta(1,2)
  \cdot\Theta(3,4) 
+\frac{(p_1+p_2)^2}{2} {\cal F}_{MN}(1,2){\cal F}^{MN}(3,4) \Biggl]\\
+&
\frac{1}{(p_1+p_4)^2}\Biggl[ \Theta(1,4) 
  \cdot\Theta(3,2)
+\frac{(p_1+p_4)^2}{2} {\cal F}_{MN}(1,4){\cal F}^{MN}(3,2) \Biggl],\\
\end{split}
\ee
where
\bsp\label{tf}
&\Theta_M(1,2) \equiv 2 p_2 \cdot \e(p_1) \,\e_M(p_2) -  2 p_1 \cdot \e(p_2)
\,\e_M(p_1)+(p_1-p_2)_M\, \e(p_1) \cdot \e(p_2),\\
&{\cal F}_{MN}(1,2) \equiv \e_M(p_1) \e_N(p_2) -  \e_N(p_1) \e_M(p_2),
\end{split}
\ee
and where $M$ and $N$ are 4-dimensional (for the ${\cal N}=2$ case), and
10-dimensional (for the ${\cal N}=8$ case) indices for which
\be
\e^M(p) =\begin{cases} (\e^\m(p), 0,\ldots,0),\quad a^\dag(p)\\
(\underbrace{0,\ldots,0}_{i+1},1,0,\ldots,0),\quad a^\dag_i(p)
\end{cases}
\quad p^M = (p^\m,0,\ldots,0).
\ee
%

\subsection{Manifestly $SO(\mathcal{N})$ invariant forms for the amplitudes}\label{so8}

Using the spinor formalism developed in section \ref{sec:onshell}, we
find that the amplitudes may be presented in a way which shows
manifest $SO({\cal N})$ invariance. Using (\ref{together}), we find
the following expressions
\bsp\label{son1}
&{\cal C}\bigl( a_{A_1} a_{A_2} a_{A_3} a_{A_4} \bigr)\\
&\qquad= -2\d_{A_1A_2}\d_{A_3A_4}\frac{\la 13\ra \la24\ra}{\la12\ra\la34\ra}
+ 2\d_{A_1A_3}\d_{A_2A_4}
+ 2\d_{A_1A_4}\d_{A_2A_3}\frac{\la 13\ra \la24\ra}{\la23\ra\la41\ra},\\
&{\cal C}\bigl( \l_{A_1} \l_{A_2}\l_{A_3} \l_{A_4} \bigr)\\
&\qquad= 2\d_{A_1A_2}\d_{A_3A_4}\frac{\la 13\ra^2 \la23\ra}{\la12\ra^2\la41\ra}
 -2\d_{A_1A_3}\d_{A_2A_4}\frac{\la34\ra}{\la12\ra}
 -2\d_{A_1A_4}\d_{A_2A_3}\frac{\la 13\ra^2 \la12\ra}{\la14\ra^2\la34\ra},
\end{split}
\ee
and, defining $\r^{A_1A_2} \equiv \frac{1}{2}\Bigl((\r^{A_1})^T
\r^{A_2}-(\r^{A_2})^T \r^{A_1}\Bigr)$, 
\bsp\label{son2}
&{\cal C}\bigl( a_{A_1} a_{A_2} \l_{A_3} \l_{A_4} \bigr)
= -\d_{A_1A_2}\d_{A_3A_4}\frac{\la 13\ra^2}{\la12\ra^2}\left(\frac{\la
  13\ra}{\la14\ra}+ \frac{\la 23\ra}{\la24\ra}\right)
+ \Bigl(\r^{A_1A_2}\Bigr)_{A_3A_4}\frac{\la31\ra}{\la14\ra},\\
&{\cal C}\bigl( a_{A_1} \l_{A_2} \l_{A_3} a_{A_4} \bigr)
=-\d_{A_1A_4}\d_{A_2A_3} \frac{\la 42\ra^2}{\la41\ra^2}\left(\frac{\la
  42\ra}{\la43\ra}+ \frac{\la 12\ra}{\la13\ra}\right)
-\Bigl(\r^{A_1A_4}\Bigr)_{A_2A_3}\frac{\la24\ra}{\la43\ra},\\
&{\cal C}\bigl( a_{A_1} \l_{A_2} a_{A_3} \l_{A_4} \bigr)
= -\d_{A_1A_3}\d_{A_2A_4}\left(\frac{\la
  12\ra}{\la14\ra}- \frac{\la 23\ra}{\la34\ra}\right)\\
&\qquad\qquad\qquad\qquad\qquad\qquad\qquad\qquad\qquad\qquad
+\Bigl(\r^{A_1A_3}\Bigr)_{A_2A_4}\left(\frac{\la12\ra}{\la14\ra}
+\frac{\la23\ra}{\la34\ra}\right).\\
\end{split}
\ee
Note that the mixed amplitudes with a fremion in the first position
can be obtained straightforwardly using (\ref{rear}) and
(\ref{fermfirst}).  The invariant expressions (\ref{son1})
and (\ref{son2}) can straightforwardly be checked to satisfy the
associated supersymmetry algebra given in (\ref{susyn28}). As an
explicit example, we consider the ${\cal N}=2$ case in the next
subsection.

\subsection{Reconstruction of helicity}

Using the ${\cal N}=2$ theory as an example, we show how the $d=4$ MHV
amplitudes are recovered. Following section \ref{sec:rech}, one may define a
three-dimensional analogue of helicity
\bsp
&a_{\pm} = \frac{1}{\sqrt{2}} \left( a_1 \pm i a_2\right),\\
&\l_{\pm} = \frac{1}{\sqrt{2}} \left( \l_1 \pm i \l_2\right),
\end{split}
\ee
from which one finds ${\cal C}(\phi_+\phi_+\phi_+\phi_+) = {\cal C}(
\phi_-\phi_-\phi_-\phi_- )=0$ for all fields. The non-zero amplitudes are as follows
\be
{\cal C}_{\text{MHV}}(a a a a) = 2\frac{\la i j\ra^4}{\la12\ra\la23\ra\la34\ra\la41\ra},\label{mhv3}
\ee
where $i,j$ denote the positions of the negative (assuming the other
two are positive) or positive (assuming the other two are negative)
helicity states. This is of course the Parke-Taylor formula. One also finds
\bsp
&{\cal C}(\l_+\l_+\l_-\l_-) =\frac{\la13\ra}{\la24\ra} {\cal
  C}(a_+ a_+ a_- a_-),\\
&{\cal C}(\l_+ \l_- \l_+ \l_-) =\frac{\la24\ra}{\la13\ra} {\cal
  C}(a_+ a_- a_+ a_-),\\
&{\cal C}(\l_+ \l_- \l_- \l_+) = \frac{\la13\ra}{\la24\ra} {\cal
  C}(a_+ a_- a_- a_+),
\end{split}
\ee
in agreement with (\ref{ffff}). The non-zero mixed fermion-boson
amplitudes are in agreement with (\ref{bbff}).

\section{Comments on loop corrections for $\mathcal{N}=8$}
\label{sec:oneloop}

We briefly comment on the integrands corresponding to the one-loop
correction to the tree-level amplitudes using the case of
$\mathcal{N}=8$ theory as an example.  The method of unitarity cuts
allows us to efficiently evaluate the one-loop contribution to the
four-particle $\mathcal{N}=8$ amplitudes from the knowledge of the
corresponding tree-level quantities. Furthermore, since the tree-level
amplitudes are nothing but the four-dimensional ones evaluated in a
boosted frame where $k_3=0$ (c.f. section \ref{dimredhelicity}), the
integrands contributing to the loop corrections to any amplitude can
be easily constructed from the known results in four dimensions. For,
instance, consider a four-particle amplitude $\mathcal{M}(1^{h_1},
2^{h_2}, 3^{h_3}, 4^{h_4})$. In $\mathcal{N}=4$ SYM in $d=4$, the
contribution from the $s$ or $t$ channel cut to this amplitude is
generically of the form
\bsp \mathcal{M}(1^{h_1}, 2^{h_2}, 3^{h_3}, 4^{h_4})|_{cut} =
&\sum_{h,h'}\int \frac{d^4k}{(2\pi)^4} 2\pi
\delta^+(l_1^2)\delta^+(l_2^2)\\ &\times
\mathcal{M}^{tree}_1(-l_1^{h}, 1^{h_1}, 2^{h_2}, l_2^{h'})
\mathcal{M}^{tree}_2(-l_2^{\bar h'}, 3^{h_3}, 4^{h_4}, l_1^{\bar h}),
\end{split}
\ee
where $\mathcal{M}_i$ are the amplitudes that contribute to the
particular cut, $l_1 = k$, and $l_2 = k - p_1 - p_2$, where $k$ is the
loop-momentum.  One replaces the delta functions by $\frac{i}{2\pi
l_{i = 1,2}^2}$to construct the full integrand of the Feynman integral
contributing to the one-loop correction to $\mathcal{M}$. Given the
relations between the amplitudes of the three and four-dimensional
theories, it readily follows that the corresponding three-dimensional
amplitude is given by 
\bsp \mathcal{M}_{d=3}(1^{h_1}, 2^{h_2},
3^{h_3}, 4^{h_4})|_{cut} = &\sum_{h,h'}\int \frac{d^3k}{(2\pi)^3} 2\pi
\delta^+(l_1^2)\delta^+(l_2^2)\\ &\times
\mathcal{M}^{tree}_{d=3,1}(-l_1^{h}, 1^{h_1}, 2^{h_2}, l_2^{h'})
\mathcal{M}^{tree}_{d=3,2}(-l_2^{\bar h'}, 3^{h_3}, 4^{h_4}, l_1^{\bar
h}),
\end{split}
\ee
where, $\mathcal{M}_{d=3}$ are obtained setting the fourth components
of all momenta to zero in the corresponding four-dimensional
quantity. In the three-dimensional context, $h_i$ corresponds to
$U(1)$ charge carried by the states.  All the algebraic identities
between various spinor products that are used to bring the integrands
of the one-loop amplitudes in $\mathcal{N}=4$ SYM to a scalar-box
integral continue to hold after the dimensional reduction to $d=3$ as
well.

Since there is only one-independent four-particle amplitude - the rest
are related to any given amplitude by the constraints of
supersymmetry - we only give the answer for the one-loop ``MHV''
amplitude in three dimensions. After accounting for the $t$ and $s$
channel cuts, one has
\be\label{3d1loop} \langle a_+a_+a_-a_- \rangle_1 = -st\langle
a_+a_+a_-a_- \rangle_0 I \ee The subscripts (0,1) refer to tree-level
and one-loop respectively, while $I$ is the three-dimensional massless
scalar box integral 
\be I = \int \frac{d^3 q}{(2\pi)^3}
\frac{1}{q^2(q+p_1)^2(q+p_1+p_2)^2(q-p_4)^2}. 
\ee
The massless scalar box integral is IR divergent in dimensions $d \leq
4$, and so is far away from convergence in $d=3$. A potential method
for regulating it is to use the Coulomb branch as has been done for
${\cal N}=4$ SYM in $d=4$ in \cite{Alday:2009zm}. We leave this issue
to a further publication. 

Although we concentrated on the ${\cal N}=8$ theory here, the $d=3$
theories with lower SUSY also share the property that the one-loop
integrands can be gotten straightforwardly through dimensional
reduction of the appropriate theory in four dimensions, and therefore
benefit from the application of unitarity-based methods employed there.  

\section{Relation to SCS theories: D2 vs. M2-brane $S$-matrices}
\label{sec:scs}

In this section, we comment on the realization of the
$SO(\mathcal{N})$ symmetric on-shell supersymmetry algebra for
$\mathcal{N}\geq 4$ SCS theories. In particular we focus on the
$\mathcal{N}=8$ BLG theory \cite{blg1,blg2,blg3,blg4}, to which the SYM
theory with $SU(2)$ gauge group is expected to flow in the deep IR. Me
make an explicit identification between the on-shell degrees of
freedom of the two theories and show that the ``matrix'' part of the
four-particle scattering matrix $S_{ijkl}(\{\mathcal{W}\}; t,s)$ of
(\ref{univ4}) is the same for both theories. Generally, one expects
the $SO(8)$ symmetry to be manifest for all observables of the SYM
theory only at $g^2_{YM} \rightarrow \infty$. However, since we have
determined $R$-symmetry structure of the four-particle amplitude to all
orders in perturbation theory, we are able to compare the manifestly
$SO(8)$ symmetric observables in both the theories in a transparent
manner.

In the case of supersymmetric Chern-Simons theories, the construction
of Gaiotto and Witten \cite{gw} and its
generalizations \cite{hlllp1,hlllp2} allow one to construct
$\mathcal{N}\geq 4$ SCS theories in a unified manner. One starts with
a symplectic group $Sp(2n)$, which contains the gauge group $G$ as a
subgroup. $Sp(2n)$ has an antisymmetric form $\omega_{AB}$ and a
Cartan metric $k^{mn}$. The generators of the gauge group $t^{mA}_B$
are $2n\times 2n$ matrices for each value of $m$. $m$ should be
regarded as an ``adjoint'' index, while $A, B$ can be thought of as
``fundamental'' indices. The gauge potential $A^m_\mu$ has a Lorentz
and an adjoint index as expected. The matter fields (bosons)
$q^A_\beta$ and (fermions) $\psi^A_{\dot \beta}$ carry two different
$SU(2)$ (dotted and undotted) indices, apart from the index
corresponding to gauge group $A$. The matter fields are taken to
satisfy the reality conditions
\be q^\dagger_{A\alpha} = \epsilon_{\alpha \beta}\omega
_{AB}q^B_\beta, \hspace{.3cm} \psi^\dagger_{A\dot \alpha} =
\epsilon_{\dot \alpha \dot \beta}\omega _{AB}\psi^B_{\dot \beta} .
\ee
The conditions for $\mathcal{N}=4$ SUSY were derived by Gaiotto and
Witten to be 
\be k_{mn} t^m_{a(b}t^n_{CD)}= 0 ,
\ee 
where the brackets
denote symmetrization of the indices.  The supersymmetry generators
act on the asymptotic states as follows \cite{scs1}
\be \mathcal{Q}_{a \alpha
\dot \beta} |q_\gamma\rangle = u_a \epsilon_{\alpha \gamma}|\psi_{\dot
\beta}\rangle, \hspace{.3cm} \mathcal{Q}_{a\alpha \dot \beta}
|\psi_{\dot \gamma}\rangle = u_a \epsilon_{\dot \beta \dot
\gamma}|q_{\alpha }\rangle. \ee The generators close as \be
\{\mathcal{Q}_{a \alpha \dot \beta}, \mathcal{Q}_{b \gamma \dot
\delta}\} = \epsilon _{\alpha \gamma}\epsilon_{\dot \beta \dot
\delta}P_{ab} .
\ee
One can add more matter multiplets, the so called twisted
hypermultiplets, (bosons) $\tilde q^A_{\dot \alpha}$ and (fermions)
$\tilde \psi ^A_\alpha$, transforming under a representation of the
gauge group generated by $\tilde t^m_{AB}$ which in general $\neq
t^m_{AB}$. The twisting refers to the interchanging of the $SU(2)$
indices for the bose and fermi particles with respect to the original
matter fields. At this level of generality one can only have
$\mathcal{N}=4$ SUSY.  It was observed in \cite{hlllp1,hlllp2} that, if
$t = \tilde t$, then one has an enhancement of SUSY to
$\mathcal{N}=5$. Furthermore if $t = \tilde t$ can be decomposed as
$(\mathcal{R}, \bar{\mathcal{R})}$, e.g. as in bifundamentals
($N,\bar{N}$) of $SU(N)$, then one has enhancement to $\mathcal{N}=6$
SUSY. Finally, if the representations are real, $\mathcal{R} =
\bar{\mathcal{R}}$, then one recovers $\mathcal{N}=8$
supersymmetry. For the last case, the only known example is
the $SU(2)$ superconformal Chern-Simons theory of BLG; we can make an
explicit identification between the degrees of freedom of the
Yang-Mills theory and the $\mathcal{N}=8$ SCS theory it is expected to
be described by at infinite coupling. In this case, one has eight real
scalars $X^{A(I)}$.  The $SU(2)$ indices of the preceding discussion
have been promoted to an $SO(8)$ index $I$. The supersymmetry
variation of the scalars is given by
\be \delta X^{A(I)} =
i\bar{\epsilon}\Gamma ^{I+2} \Psi ^A .
\ee
For the SYM theory, the
variation of the seven scalars obtained by the dimensional reduction
of the $\mathcal{N}=1, d=10$ theory $\Phi^{I= 3\cdots 9}$ is given in
the three-dimensional notation by 
\be \delta \Phi^{A(I)} =
i\bar{\epsilon}\Gamma ^{I} \Psi ^A .
\ee 
The ``gauge'' indices of the scalars of the BLG theory correspond to a
3-algebra (which for the BLG theory is an $SU(2)$ algebra in
disguise \cite{raamsdonk}), while for the SYM theory they are the
adjoint $SU(2)$ indices. However, since we are only interested in the
color-ordered amplitudes, we can disregard the gauge index and
immediately see that the on-shell supersymmetry variations of the seven
scalars of the SYM theory coincide with the variations of $X^{I = 1
\cdots 7}$. However,
\be \delta X^{A(8)} =
i\bar{\epsilon}\Gamma ^{11} \Psi ^A = i\bar{\epsilon} \Psi ^A,
\ee 
where we have used the $d=10$ Weyl condition.  The on-shell version of
this transformation is
\be (Q_A)_\alpha|X^8\rangle =
\frac{1}{2}u_\alpha |\lambda_A\rangle ,
\ee 
which, as we have seen previously is exactly the same as the
transformation law for $a_1$, the scalar that is dual to the gauge
field of the Yang-Mills theory.  Thus, for the on-shell supersymmetry
algebra $X^I_{BLG} \leftrightarrow (\Phi^I, A_\mu)_{SYM}$. The rest of
the on-shell SUSY algebra is uniquely determined by the requirements
of the closure of the generators on to $P$. Thus, the full on-shell
algebras of the $\mathcal{N}=8$ SCS and SYM theories are the
same. This immediately implies that the ``matrix'' structures of the
$S$-matrices of the two theories are also
identical. $S_{ijkl}(\{\mathcal{W}\}; t,s)$ for the SYM theory can be
readily extracted from the expressions in section \ref{so8} which is
the same as the corresponding quantity for the SCS theory found in
\cite{scs1}. Thus the flow of the D2-brane theory to the M2-brane
model (at the level of four-particle amplitudes) corresponds to
understanding how the one independent amplitude which takes on the MHV
form (\ref{mhv3}) at extreme weak coupling, flows to the four-boson
scattering amplitude given in \cite{scs1}, at infinite
coupling. Obviously, obtaining the full interpolating form of the
four-particle amplitude would be an enormous progress towards
establishing the connection between D2 and M2-brane theories.

\section{Concluding remarks}\label{sec:conc}

We have presented an on-shell formalism that reveals several algebraic
properties of $S$-matrices of three-dimensional Yang-Mills theories
that are not evident at the level of the corresponding Lagrangians. In
particular our framework uncovers a hidden $U(1)$ symmetry, which is
an augmentation of the helicity degree of freedom of four-dimensional
parent theories, to a continuous symmetry upon dimensional reduction.
This emergent $U(1)$ lifts the $SO(\mathcal{N}-1)$ symmetry of the
gauge theory Lagrangians to an $SO(\mathcal{N})$ symmetry of the
on-shell algebra and $S$-matrix. We have been able to confirm that the
manifest $SO(\mathcal{N})$ invariance is indeed realized by four-particle
amplitudes to all orders in perturbation theory while presenting
arguments in favor of the same phenomenon for higher-particle
amplitudes. As an application of the methodology presented in this
work, we used the manifest $SO(8)$ invariance of the four-particle
amplitude of the $\mathcal{N}=8$ theory to show that the amplitude of
the SYM theory is the same as that of the BLG theory up to a single
function. We have also presented detailed results for the tree-level
forms of the four-particle amplitudes of all the gauge theories
considered in this paper while paying special attention to the
$SO(\mathcal{N})$ invariance of the results. Other than the issue of
symmetry enhancement, we have also shown that recursion relations for
$d=3$ SYM theories can readily be obtained via dimensional reduction.

The results of this paper point to several exciting directions for
future research. At tree-level, it is a simple exercise to obtain
three-dimensional results via dimensional reduction from four
dimensions. Indeed, even at the loop-level, it would appear that the
$d=3$ integrands are trivial generalizations of their $d=4$
counterparts, and in this sense are determined by the structure of the
parent theory, e.g. for ${\cal N}=4$ in $d=4$, this structure
is believed to be captured by a Grassmannian \cite{ArkaniHamed:2009dn}.
The structure of IR divergences, however, may be very
different in three dimensions. The question of how the various
symmetries and formulations of four-dimensional amplitudes translate
to three dimensions once internal momenta have been integrated over is
a very interesting question, and one which we hope to report on in the
near future. 

Finally, it would be extremely interesting to build on the connection
between the $S$-matrices of M2 and D2-brane theories pointed out in
this paper. At the level of four-particle amplitudes, we have reduced the
problem of understanding the flow of $\mathcal{N}=8$ SYM theory to the
BLG theory to understanding the asymptotic behavior of a single
function. Given the recent hints of the existence of a twistorial
structure and Yangian symmetries for the scattering matrices of SCS
theories \cite{scs2,scs3,scs4} it is perhaps not inconceivable that the
relationship between the SCS and SYM theories can be understood very
concretely at the on-shell level by uncovering the corresponding
algebraic structures for the SYM amplitudes.

\section*{Acknowledgements}

We would like to thank Johannes Henn and V.P. Nair for important discussions. 

\appendix

\section{Dimensional reductions}
\label{app:dimred}

\subsection{Conventions}

We work in mostly positive signature. Our fermions are Majorana and
obey $\bar\l_A = \l_A^TC_d$, where $C_d$ is the $d$-dimensional charge
conjugation matrix which is identified with the zero component gamma
matrix. Our scalars are real. Each field in the theory $\phi = \phi^a
T^a$, $a=1,\ldots,N^2-1$, where the generators $T^a$ of $SU(N)$ are
$N\times N$ matrices obeying the following identities
\bsp
&T^a T^a = \frac{N^2-1}{2N} {\bf 1}, \quad \Tr(T^aT^b) = \frac{1}{2}
\d^{ab}, \quad [T^a,T^b] = i f^{abc} T^c, \quad f^{abc}f^{abd} =
N\d^{cd},\\
&\{T^a,T^b\} = \frac{1}{N}\d^{ab} {\bf 1} + d^{abc} T^c,
\end{split}
\ee
and the gauge covariant derivative is defined as 
\bsp
&D_\m \phi = \p_\m\phi -i[A_\m,\phi],\\ 
&F_{\m\n} = \p_\m A_\n - \p_\n A_\m -i[A_\m,A_\n].
\end{split}
\ee
The free field propagators for the various fields are given by
\bsp
&\la \Phi^a_i(p) \Phi^b_j(-p) \ra = -\d_{ij}\frac{ig^2\d^{ab}}{p^2}, \quad
\la A_\m^a(p) A_\n^b(-p) \ra = - \d^{ab}\frac{ig^2\eta_{\m\n}}{p^2},\\
&\la \l^a_{A\,\a}(p) \l^b_{B\,\b}(-p) \ra =  -\d_{AB}\d^{ab}\frac{ig^2p_\m \left(\g^\m
  C_3^{-1}\right)_{\a\b}}{p^2}.
\end{split}
\ee
We have chosen Feynman gauge for the gauge field. The ghost action is
not given as we are working at tree-level.

\subsection{Spinor identities}

Defining $\la p| \equiv \bar u(p)$ and $|p \ra \equiv u(p)$, we note
the following relations
\bsp
&\la ij\ra = -\la ji\ra,\\
&{\not \e}(p_1,p_2) u(p_3) = -2\frac{\la13\ra}{\la 12\ra} u(p_2) +
u(p_3),\\
&\bar  u(p_3){\not \e}(p_1,p_2) = -2\frac{\la32\ra}{\la 12\ra} \bar u(p_1) +
\bar u(p_3),\\
&p_3 \cdot \e(p_1,p_2) = 2\left(p_1\cdot p_3\right) \frac{\la 2
  3 \ra}{\la 2 1\ra\la 3 1 \ra},\\
&\la 1| \g^\m|2\ra\la3|\g_\m|4\ra =
\la13\ra\la42\ra+\la23\ra\la41\ra,\\
&\la12\ra\la34\ra=\la23\ra\la41\ra-\la13\ra\la42\ra,\\
&p_i\cdot p_j = - \frac{\la i j \ra^2}{2},\\
&\frac{\la14\ra}{\la23\ra}=-\frac{\la24\ra}{\la13\ra},\quad
\frac{\la12\ra}{\la34\ra}=-\frac{\la13\ra}{\la24\ra}.\\
\end{split}
\ee
The last two relations follow from momentum conservation for
4-particle scattering.

\subsection{${\cal N}=2$ theory}

We begin with ${\cal N}=1$ SYM in $d=4$, whose action is
\be\label{d4N4act}
S_{{\cal N}=1,\,d=4} = \frac{1}{g^2} \Tr \int d^4 x 
\left(-\frac{1}{2} F_{MN} F^{MN} + i \bar \Psi \G^M D_M \Psi \right),
\ee
where $M,N = 0,\ldots,3$ and we are using mostly plus signature
\be
\eta_{MN} = \text{diag}(-1,1,1,1),
\ee
and $\Psi$ ia a four-component Majorana spinor, and we use the
real representation of the gamma-matrices provided in (\ref{4gamma}).
The charge conjugation matrix is $C_4 = \G^0$ and $\bar\Psi \equiv \Psi^T
C_4$. Let us write the Majorana spinor $\Psi$ as follows
\be
\Psi = \begin{pmatrix} \l_1\\ \l_2 \end{pmatrix}.
\ee
Under dimensional reduction whereby we eliminate the last dimension,
i.e. $\p_3 \to 0$, we obtain the following action
\be
S_{{\cal N}=2,\,d=3} = \frac{1}{g^2} \Tr \int d^3 x 
\left(-\frac{1}{2} F_{MN} F^{MN} + i \bar \l_A \g^\m D_\m \l_A 
+\e_{AB}\bar \l_A [\Phi,\l_B] 
\right),
\ee
where $A=1,2$, $\m = 0,1,2$, $\e_{12} = +1$, and where the real representation of the
three-dimensional gamma matrices used is
\be
\g^\m =\left(i\s^2,\s^1,\s^3 \right), \quad C_3 = i \s^2,\quad
\eta_{\m\n} = \text{diag}(-1,1,1),
\ee
and $\bar \l \equiv \l^T C_3$. Note that we have introduced 
\be
\Phi \equiv A_3,
\ee
and assumed that $\p_3 = 0$ in the kinetic term for the gauge fields.

\subsubsection{Supersymmetry}
\label{app:N2susy}

We note the ${\cal N}=2$ SUSY transformations. Those of the
original action (\ref{d4N4act}) are
\be
\d A_N = -2i \bar \Psi \G_N \e,\qquad \d\Psi = F^{MN}\G_{MN}\, \e.
\ee
We decompose $\e$ in terms of $d=3$ SUSY parameters
\be
\e = \begin{pmatrix} \eta_1\\ \eta_2 \end{pmatrix}.
\ee
We find
\bsp
&\d A_\m = -2 i \bar \l_A \g_\m \eta_A,\\
&\d \Phi = -2i \e_{AB} \bar \l_A \eta_B,\\
&\d \l_A = F_{\m\n} \g^{\m\n} \eta_A + 2 \p_\m \Phi\, \e_{AB} \g^\m  \eta_B.
\end{split}
\ee
From the standard mode expansions
(\ref{modeexp}) we have
\be
a_1^\dag = \e_\m(p) A^\m(p),\quad a_2^\dag = \Phi(p),\quad
\l_A^\dag = \frac{1}{2p^0} u(p) \l_A(p).
\ee
Note that $u(p) u(p) = 2 p^0$.
In appendix \ref{app:sc}, the supercharge is calculated in the
four-dimensional formalism. Plugging in the mode expansions and using
\be
u(p) {\not p} = -2p^0 \bar u(p),
\ee
and
\be
\bar u {\not \e}  = -\bar u ,
\ee
which follows from (\ref{fi}), one recovers (\ref{Q1}) and (\ref{Q2}).

\subsection{${\cal N}=8$ theory}

The 10-dimensional gamma matrices, in mostly positive signature, may
be expressed using (\ref{4gamma}), as follows
\be
\wt\G^M = \g^M \otimes \mathbbm{1}_8,\quad \wt\G^I = i\g^{0123} \otimes \eta^I.
\ee
where the $SO(6)$ gamma matrices $\eta^I$ are given by
\bsp
\eta^I = \bigl\{ &\s^2 \otimes \s^2 \otimes \s^1,\,
-\s^2\otimes\s^2\otimes\s^3,\,
\s^2\otimes\mathbbm{1}\otimes\s^2,\\
-&\s^1\otimes\s^1\otimes\s^2,\,
-\s^1\otimes\s^2\otimes\mathbbm{1},\,
\s^1\otimes\s^3\otimes\s^2 \bigr\}.
\end{split}
\ee
The 6-dimensional charge conjugation matrix is
\be
C_6 = \s^1\otimes\mathbbm{1}\otimes\mathbbm{1},
\ee
while
\be
\eta^{123456} =i \s^3\otimes\mathbbm{1}\otimes\mathbbm{1}.
\ee
This gives 
\bsp
&\wt\G^{11} = \wt\G^{0123456789} = i\begin{pmatrix} &0 &\mathbbm{1}_2\\ &-\mathbbm{1}_2 &0
\end{pmatrix}\otimes \begin{pmatrix} &\mathbbm{1}_4 &0\\ &0
  &-\mathbbm{1}_4 \end{pmatrix},\\
&C_{10} =   \begin{pmatrix} &i\s^2 &0\\ &0 &-i\s^2
\end{pmatrix} \otimes \begin{pmatrix} &0 &\mathbbm{1}_4\\ 
  &\mathbbm{1}_4 &0\end{pmatrix}.
\end{split}
\ee
Implementing the Weyl and Majorana conditions
\be
\wt\G^{11} \Psi = \Psi,\qquad \Psi^\dag \wt\G^0 = \Psi^T C_{10} \equiv \bar\Psi,
\ee
one obtains
\bsp\label{decomp1}
\Psi = &\frac{1}{2} \begin{pmatrix} &i\\&0\\&1\\&0\end{pmatrix}
\otimes  \begin{pmatrix}&\chi_1+i\chi_2\\&\chi_3+i\chi_4\\&0\\&0
\end{pmatrix}
+ \frac{1}{2} \begin{pmatrix} &-i\\&0\\&1\\&0\end{pmatrix}
\otimes  \begin{pmatrix}&0\\&0\\&\chi_1-i\chi_2\\&\chi_3-i\chi_4
\end{pmatrix}\\
+&\frac{1}{2} \begin{pmatrix} &0\\&-i\\&0\\&1\end{pmatrix}
\otimes  \begin{pmatrix}&0\\&0\\&\chi_5+i\chi_6\\&\chi_7+i\chi_8
\end{pmatrix}
+\frac{1}{2} \begin{pmatrix} &0\\&i\\&0\\&1\end{pmatrix}
\otimes  \begin{pmatrix}&\chi_5-i\chi_6\\&\chi_7-i\chi_8\\&0\\&0
\end{pmatrix},
\end{split}
\ee
where the $\chi_A$ are real 2-spinors. Redefining the fields in terms
of 8 $d=3$ Majorana 2-spinors $\l_A$

\bsp\label{decomp2}
&\chi_1 = \begin{pmatrix} (\l_1)_1\\ (\l_5)_1 \end{pmatrix},\quad
\chi_5 = \begin{pmatrix} (\l_1)_2\\ (\l_5)_2 \end{pmatrix},\quad
\chi_3 = \begin{pmatrix} (\l_3)_1\\ (\l_7)_1 \end{pmatrix},\quad
\chi_7 = \begin{pmatrix} (\l_3)_2\\ (\l_7)_2 \end{pmatrix},\\
&\chi_2 = \begin{pmatrix} (\l_2)_1\\ (\l_6)_1 \end{pmatrix},\quad
\chi_6 = -\begin{pmatrix} (\l_2)_2\\ (\l_6)_2 \end{pmatrix},\quad
\chi_4 = \begin{pmatrix} (\l_4)_1\\ (\l_8)_1 \end{pmatrix},\quad
\chi_8 = -\begin{pmatrix} (\l_4)_2\\ (\l_8)_2 \end{pmatrix},\\
\end{split}
\ee
where the index outside the bracket denotes the first or second
component of the spinor, one obtains from the ${\cal N}=1$, $d=10$ action
\be
 S_{{\cal N}=1,\,d=10} = \frac{1}{g^2} \Tr \int d^{10} x 
\left(-\frac{1}{2} F_{\bar M\bar N} F^{\bar M \bar N} + i \bar \Psi
 \wt\G^{\bar M} D_{\bar M} \Psi \right),
\ee
the ${\cal N}=8$, $d=3$ action
\bsp
 S_{{\cal N}=8,\,d=3} = \frac{1}{g^2} \Tr \int d^3 x 
\Bigl(-\frac{1}{2} F_{\bar M\bar N} F^{\bar M \bar N} + i \bar \l_A
 \g^{\m} D_{\m} \l_A 
 +\r^i_{AB} \bar  \l_A [\Phi_i ,\l_B] \Bigr),
\end{split}
\ee 
where $\r^C_{AB}$ are the matrices relating the ${\bf 8}_v$
(index $C=1,\ldots,8$), ${\bf 8}_s$ (index $A=1,\ldots,8$), and  ${\bf
  8}_c$ (index $B=1,\ldots,8$) of $SO(8)$. Explicitly we have that $A_{\bar M} =
(A_\m,\Phi_i)$, $i=2,\ldots,8$ and 
\bsp\label{rhos}
\r^C_{AB} = \Bigl\{&\mathbbm{1}\otimes\mathbbm{1}\otimes\mathbbm{1},~
\mathbbm{1}\otimes\mathbbm{1} \otimes i\s^2,
-\s^1\otimes i\s^2\otimes \s^3,~
\s^3\otimes i\s^2\otimes \s^3,
-i\s^2\otimes\mathbbm{1}\otimes \s^3,~\\
i&\s^2\otimes\s^1\otimes\s^1,~
\mathbbm{1}\otimes i\s^2\otimes\s^1,~
-i\s^2\otimes\s^3\otimes\s^1 \Bigr\}.
\end{split}
\ee
One has that $\r^D_{AC} \r^E_{BC} + \r^E_{AC} \r^D_{BC} =
2\d^{DE}\d_{AB}$, $\r^{i}_{AB} = -\r^{i}_{BA}$.

The supersymmetry of the theory may be gotten by following steps
similar to those in section \ref{app:N2susy}. The supercharge
(\ref{master}) may be used (taking the ${\cal N}=1$ theory in $d=10$)
along with the decomposition given in (\ref{decomp1}) and
(\ref{decomp2}). The results may be compactly expressed as
\bsp\label{susyn28}
Q_A^\a |a_B\ra = \frac{1}{2}u^\a\,\r^B_{AC}|\l_C\ra,\qquad
Q_A^\a |\l_B\ra = -\frac{1}{2} u^\a\,\r^C_{AB} |a_C\ra.
\end{split}
\ee

\section{Supercharge for the ${\cal N}=1$ theory}
\label{app:sc}

We begin with the Lagrangian
\be
{\cal L} = -\frac{1}{2} F_{MN} F^{MN} + i \bar \Psi \G^M D_M \Psi.
\ee 
The SUSY variations of the fields are as follows
\be
\d A_N = -2i\bar\Psi \G_N \e,\qquad
\d \Psi = F_{PQ} \G^{PQ} \e.
\ee
The variation of the action is then (knowing that the interacting
theory is supersymmetric, we set the coupling to zero and so take $D_M
\to \nabla_M$ (to remain as general as possible we use the $\nabla$ in
place of the partial derivative))
\be\label{varL}
\d {\cal L} = 4i F^{MN} \nabla_M \left(\bar\Psi \G_N \e \right) + 
i\bar\Psi \G^M \nabla_M \left(F_{PQ}\G^{PQ} \e\right)
+i\,\overline{\left(F_{PQ}\G^{PQ} \e\right)} \G^M\nabla_M \Psi.
\ee
The bar operation is $\bar\Psi \equiv \Psi^TC$, where $C^T=-C$ and
$C\G^M C^{-1} = -\left(\G^M\right)^T$. Using integration by parts on
the second term in (\ref{varL}) and then reversing the order in the
last term one finds
\be
\d {\cal L} = 4i F^{MN} \nabla_M \left(\bar\Psi \G_N \e \right)  
-2i \left(\nabla_M \bar\Psi\right) \G^M F_{PQ}\G^{PQ} \e
+\nabla_M \left( i\bar\Psi \G^M \G^{PQ} F_{PQ} \e\right).
\ee
Now we use the fact that
\be\label{idenF}
\G^M\G^{PQ}F_{PQ} = 2F^{MQ}\G_Q +\G^{MPQ} F_{PQ},
\ee
to produce
\be
\d {\cal L} = 4i F^{MN} \bar\Psi \G_N \nabla_M \e   
-2i \left(\nabla_M \bar\Psi\right) \G^{MPQ} F_{PQ} \e
+\nabla_M \left( i\bar\Psi \G^M \G^{PQ} F_{PQ} \e\right).
\ee
Now notice that $\G^{MPQ}\nabla_M F_{PQ}$ is identically zero since we
have antisymmetrized double partial derivatives acting on the gauge
field in the field strength. Thus, integration by parts on the middle
term above produces
\bsp
\d {\cal L} = &4i F^{MN} \bar\Psi \G_N \nabla_M \e   
+2i  \bar\Psi \G^{MPQ} F_{PQ} \nabla_M\e\\
&+\nabla_M\left(-2i  \bar\Psi \G^{MPQ} F_{PQ} \e\right)
+\nabla_M \left( i\bar\Psi \G^M \G^{PQ} F_{PQ} \e\right)\\
&=2i  \bar\Psi \G^M\G^{PQ} F_{PQ} \nabla_M\e
+\nabla_M\left( i\bar\Psi\left( \G^M\G^{PQ} - 2\G^{MPQ}\right)F_{PQ}\e\right),
\end{split}
\ee
where we have made use of (\ref{idenF}) in the second equality. In
flat space $\nabla_M\e=0$ and we can build the conserved Noether current
associated with the symmetry as usual
\be
j^M = \frac{\d {\cal L}}{\d(\p_MA_N)} \d A_N +  \frac{\d {\cal
    L}}{\d(\p_M\Psi)} \d \Psi - {\cal J}^M ,
\ee
where ${\cal J}^M$ is the total derivative arising from the variation
of the action, i.e. $\d{\cal L} = \nabla_M {\cal J}^M$. We therefore
find 
\bsp
j^M &= 4iF^{MN} \bar\Psi \G_N \e + i\bar\Psi\G^M \G^{PQ} \e F_{PQ}
- \left( i\bar\Psi\left( \G^M\G^{PQ} -
    2\G^{MPQ}\right)F_{PQ}\e\right)\\
&=  4iF^{MN} \bar\Psi \G_N \e +2i \bar\Psi \G^{MPQ} F_{PQ} \e\\
&=2i\bar\Psi\G^M\G^{PQ} F_{PQ}\e
\end{split}
\ee
where we have made use of (\ref{idenF}) in the last equality. The
supercharge is then given by
\be\label{master}
Q = \int_{\text{space}} j^0 = 2i  \int_{\text{space}} \bar\Psi \G^0\G^{PQ}F_{PQ}\e.
\ee

\bibliography{scattering}
\end{document}